\documentclass[twoside,leqno,twocolumn]{article}

\usepackage{ltexpprt}
\usepackage{enumerate}
\usepackage{epsfig}
\usepackage{graphicx}
\usepackage{amsmath}
\usepackage{amssymb}
\usepackage{algorithm}
\usepackage{url}
\usepackage{caption}
\usepackage{subcaption}
\usepackage{wrapfig}
\usepackage{paralist}
\usepackage{xcolor}
\usepackage{booktabs} % for table's toprule, midrule, etc.
\usepackage{grffile}

\begin{document}
\title{FairPlay: Fraud and Malware Detection in Google Play}
\author{
Mahmudur Rahman\\
Florida Int'l Univ.\\
mrahm004@fiu.edu
\and
Mizanur Rahman\\
Florida Int'l Univ.\\
mrahm031@fiu.edu
\and
Bogdan Carbunar\\
Florida Int'l Univ.\\
carbunar@gmail.com
\and
Duen Horng Chau\\
Georgia Tech\\
polo@gatech.edu
}

\date{}

\maketitle

\begin{abstract}
Fraudulent behaviors in Google's Android app market fuel search rank abuse and
malware proliferation.  We present FairPlay, a novel system that uncovers both
malware and search rank fraud apps, by picking out trails that fraudsters leave
behind.  
To identify suspicious apps, FairPlay's PCF algorithm correlates review activities and uniquely
combines detected review relations with linguistic and behavioral signals
gleaned from longitudinal Google Play app data.
We contribute a new longitudinal app dataset to the community, which consists
of over 87K apps, 2.9M reviews, and 2.4M reviewers, collected over half a year.
FairPlay achieves over 95\% accuracy in classifying gold standard datasets of
malware, fraudulent and legitimate apps.  We show that 75\% of the identified
malware apps engage in search rank fraud.  FairPlay discovers hundreds of
fraudulent apps that currently evade Google Bouncer's detection technology,
and reveals a new type of attack campaign, where  users are harassed into
writing positive reviews, and install and review other apps.
\end{abstract}

\vspace{5pt}
\section{Introduction}

The commercial success of Android app markets such as Google
Play~\cite{GooglePlay} has made them a lucrative medium for committing fraud
and malice.  Some fraudulent developers deceptively boost the search ranks and
popularity of their apps (e.g., through fake reviews and bogus installation
counts)~\cite{GPlay.Fake}, while malicious developers use app markets as a
launch pad for their
malware~\cite{Malware.PCWorld,Malware.PCMag,Malware.Fortune,Malware.Forbes}.
%
%The motivation for such behaviors is impact, as increased popularity leads to
%financial benefits and simplifies malware proliferation.

Existing mobile malware detection solutions have limitations. For instance,
while Google Play uses the Bouncer system~\cite{OM12} to remove malware, out of
the $7,756$ Google Play apps we analyzed using
VirusTotal~\cite{VirusTotal}, $12$\% ($948$) were flagged by at least one
anti-virus tool and $2$\% ($150$) were identified as malware by at least $10$
tools (see Figure~\ref{fig:malware.apps}).  Previous work has focused on
dynamic analysis of app executables~\cite{BZNT11,SKEGW12,GZZZJ12} as well as
static analysis of code and permissions~\cite{SLGPNRM12,PGSLQPNRM12,YSM14}.
However, recent Android malware analysis revealed that malware evolves quickly to bypass anti-virus tools~\cite{ZJ12}.
%Zhou and Jiang~\cite{ZJ12}'s
%$1,200$ 
%samples has

\begin{figure}
\centering
\includegraphics[width=0.45\textwidth]{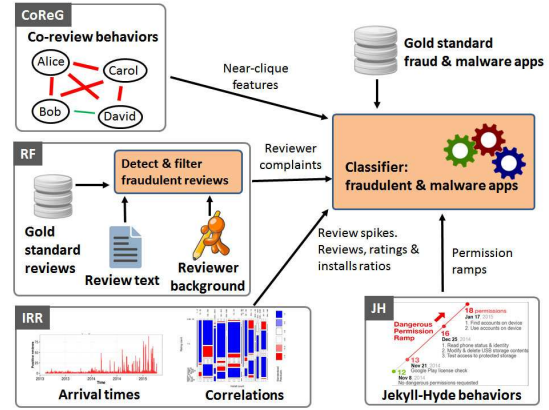}
\vspace{-5pt}
\caption{FairPlay system architecture. The CoReG module identifies suspicious,
time related co-review behaviors. The RF module uses linguistic tools to detect
suspicious behaviors reported by genuine reviews. The IRR module uses
behavioral information to detect suspicious apps.  The JH module identifies
permission ramps to pinpoint possible Jekyll-Hyde app transitions.
\label{fig:fair:play}}
\vspace{-35pt}
\end{figure}

In this paper, we seek to identify both malware and search rank fraud targets in
Google Play. This combination is not arbitrary: we posit that malicious
developers resort to search rank fraud to boost the impact of their malware.

Unlike existing solutions, we build this work on our observation that
fraudulent and malicious behaviors leave behind telltale signs on app markets.
We uncover these nefarious acts by picking out such trails.
%traces distinguishable from those of benign behaviors. 
For instance, the high cost of setting up valid
Google Play accounts forces fraudsters to reuse their accounts across review
writing jobs, making them likely to review more apps in common than regular
users. Resource constraints can compel fraudsters to post reviews within
short time intervals. Legitimate users affected by malware may report
unpleasant experiences in their reviews. Ramps in the number of ``dangerous''
permissions requested by apps may indicate benign to malware (Jekyll-Hyde)
transitions.

\noindent
{\bf Contributions and Results}.
We propose FairPlay, a system that leverages the above observations to
efficiently detect Google Play fraud and malware (see
Figure~\ref{fig:fair:play}). Our major contributions are:

\noindent
{\small $\bullet$}
{\bf A unified relational, linguistic and behavioral approach}.
We formulate the notion of  {\it co-review graphs} to model
 reviewing relations between users.  We develop PCF, an efficient
algorithm to identify temporally constrained, co-review  pseudo cliques ---
formed by reviewers with substantially overlapping co-reviewing activities across short time windows. 
We use linguistic and behavioral information to
(i) detect genuine reviews from which we then 
(ii) extract user-identified fraud and malware indicators.
%
%\ratul{from reviews detected as genuine by classifiers.}
%
In addition, we detect apps with (i) permission request ramps, (ii)
``unbalanced'' review, rating and install counts, and (iii) suspicious review
spikes. We generate $28$ features, and use them to train supervised learning
algorithms [$\S$~\ref{sec:fair:play}].

\noindent
{\small $\bullet$}
{\bf Novel longitudinal and gold standard datasets}.
We contributed a longitudinal dataset of $87,223$ freshly posted Google Play apps
(along with their $2.9$M reviews, from $2.3$M reviewers) collected between
October 2014 and May 2015.  
We have leveraged search rank fraud expert contacts
in Freelancer~\cite{Freelancer}, anti-virus tools and manual verifications to
collect gold standard datasets of hundreds of fraudulent, malware and
benign apps. 
We will publish these datasets alongside this work
[$\S$~\ref{sec:data}].

\noindent
{\small $\bullet$}
{\bf High Accuracy}.
FairPlay achieves over $97$\% accuracy in classifying fraudulent
and benign apps, and over $95$\% accuracy in classifying malware and benign
apps.  FairPlay significantly outperforms the malware indicators of Sarma et
al.~\cite{SLGPNRM12}. 
Furthermore, we show that malware often engages in search
rank fraud as well: When trained on fraudulent and benign apps, FairPlay
flagged as fraudulent more than $75$\% of the gold standard malware apps
[$\S$~\ref{sec:evaluation:app}].

%This confirms our conjectures that clues gleaned from app markets can pinpoint
%suspicious apps, and that malicious developers use search rank fraud to promote
%malware.

\noindent
{\small $\bullet$}
{\bf Real-world Impact: Uncover Fraud \& Attacks}.
FairPlay discovers hundreds of fraudulent apps that currently evade Google
Bouncer's detection technology. We show that these apps are indeed suspicious:
the reviewers of $93.3$\% of them form at least $1$ pseudo clique and $55$\% of
these apps have at least $33$\% of their reviewers involved in a pseudo clique.
In addition, FairPlay enabled us to discover a novel, {\it coercive campaign}
attack type, where app users are harassed into writing a positive review for
the app, and install and review other apps [$\S$~\ref{sec:evaluation:field} \&
$\S$~\ref{sec:evaluation:coercive}].

\section{Background, Related Work,\\ and Our Differences}
\label{sec:background}

%
% for journal
%
%\begin{figure}
%\centering
%\includegraphics[width=3.3in]{./figures/google.play/google.play.eps}
%\caption{Google Play participants and relations.}
%\label{fig:google.play}
%\end{figure}

\noindent
{\bf System model}.
We focus on the Android app market ecosystem of Google Play. The participants,
consisting of users and developers, have Google accounts. Developers create and
upload apps, that consist of executables (i.e., ``apks''), a set of required
permissions, and a description. The app market publishes this information,
along with the app's received reviews ($1$-$5$ stars rating \& text), ratings
($1$-$5$ stars, no text), aggregate rating (over both reviews and ratings),
install count range (predefined buckets, e.g., $50$-$100$, $100$-$500$), size,
version number, price, time of last update, and a list of ``similar'' apps.

%The text is optional and consists of a title
%and a description.
%Google Play limits the number of reviews displayed for an
%app to $4,000$.
%
%Google Play publishes the total number of reviews received and their aggregate
%rating. It also publishes the app's install range, consisting of
%predefined buckets (e.g., $50$-$100$, $100$-$500$).  Reviewers may have Google+
%accounts, in which case they have followers.

%Figure~\ref{fig:google.play} illustrates the participants in Google Play and
%their relations.

%
% for journal
%
%\begin{figure}
%\centering
%\vspace{-15pt}
%\includegraphics[width=0.49\textwidth]{./figures/install_job_large.eps}
%\caption{An ``install job'' posting from Freelancer~\cite{Freelancer}, asking
%for 2000 installs within 3 days (in orange), in an organized way that includes
%expertise verifications and provides secrecy assurances (in blue). Text has
%been enlarged for easier reading.}
%\label{fig:job:install}
%\vspace{-25pt}
%\end{figure}

\noindent
{\bf Adversarial model}.
We consider not only malicious developers, who upload malware, but also
rational fraudulent developers. Fraudulent developers attempt to tamper with
the search rank of their apps. While Google keeps secret the criteria used to
rank apps, the reviews, ratings and install counts are known to play a
fundamental part (see e.g.,~\cite{Ankit.Jain}. Fraudulent developers
often rely on crowdsourcing sites~\cite{Freelancer,Fiverr,BestAppPromotion} to
hire teams of workers to commit fraud collectively.
%
%emulating realistic, spontaneous activities from unrelated people (i.e.,
%``crowdturfing''~\cite{WWZZMZZ12}).

To review or rate an app, a user needs to have a Google account, register a
mobile device with that account, and install the app on the device. This
process complicates the job of fraudsters, who are thus more likely to reuse
accounts across review writing jobs.

\subsection{Research in Android Malware Detection.}
\label{sec:related:malware}

%Zhou and Jiang~\cite{ZJ12} collected and characterized $1,200$ Android malware
%samples, and reported the ability of malware to quickly evolve and bypass the
%detection mechanisms of anti-virus tools.
%
Burguera et al.~\cite{BZNT11} used crowdsourcing to collect system call traces
from real users, then used a ``partitional'' clustering algorithm to classify
benign and malicious apps.  Shabtai et al.~\cite{SKEGW12} extracted
features from monitored apps (e.g., CPU consumption, packets sent, running
processes) and used machine learning to identify malicious apps. Grace et
al.~\cite{GZZZJ12} used static analysis to efficiently identify high and medium
risk apps.

Previous work has also used app permissions to pinpoint
malware~\cite{SLGPNRM12,PGSLQPNRM12,YSM14}. Sarma et al.~\cite{SLGPNRM12} use
risk signals extracted from app permissions, e.g., rare critical permissions
(RCP) and rare pairs of critical permissions (RPCP), to train SVM and inform
users of the risks vs. benefits tradeoffs of apps. In
$\S$~\ref{sec:evaluation:app} we use Sarma et al.~\cite{SLGPNRM12}'s solution
as a baseline, and show that FairPlay significantly improves on its
performance.

Peng et al.~\cite{PGSLQPNRM12} propose a score to measure the risk of apps,
based on probabilistic generative models such as Naive Bayes.  Yerima et
al.~\cite{YSM14} also use features extracted from app permissions, API calls
and commands extracted from the app executables.

%Sahs and Khan~\cite{SK12} used features extracted from app permissions and
%control flow graphs to train an SVM classifier on 2000 benign and less than
%100 malicious apps.  Sanz et al.~\cite{SSLUBA13} rely strictly on permissions
%as sources of features for several machine learning tools. They use a dataset
%of around 300 legitimate and 300 malware pps.

%Google has deployed Bouncer, a framework that monitors published apps to detect
%and remove malware. Oberheide and Miller~\cite{OM12} have analyzed and revealed
%details of Bouncer (e.g., based in QEMU, using both static and dynamic
%analysis). Bouncer is not sufficient - our results show that 948 apps out of
%7,756 apps that we downloaded from Google Play are detected as suspicious by at
%least 1 anti-virus tool.

Instead of analyzing app executables, FairPlay employs a unified relational,
linguistic and behavioral approach based on longitudinal app data. FairPlay's
use of app permissions differs from existing work through its focus on the
temporal dimension, e.g., changes in the number of requested permissions, in
particular the ``dangerous'' ones. We observe that FairPlay identifies and
exploits a new relationship between malware and search rank fraud.

\subsection{Research on Graph Based Opinion Spam Detection.}
\label{sec:related:spam}

Graph based approaches have been proposed to tackle opinion
spam~\cite{YL15,ACF13}.  Ye and Akoglu~\cite{YL15} quantify the chance of a
product to be a spam campaign target, then cluster spammers on a 2-hop subgraph
induced by the products with the highest chance values. Akoglu et
al.~\cite{ACF13} frame the fraud detection as a signed network classification
problem and classify users and products, that form a bipartite network,
using a propagation-based algorithm.

FairPlay's relational approach differs as it identifies apps reviewed in a
contiguous time interval, by groups of users with a history of reviewing apps
in common. FairPlay combines the results of this approach with behavioral and
linguistic clues, extracted from longitudinal app data, to detect both search
rank fraud and malware apps. We emphasize that search rank fraud goes beyond
opinion spam, as it implies fabricating not only reviews, but also user app
install events and ratings.

\section{The Data}
\label{sec:data}

We have collected longitudinal data from $87$K+ newly released apps over more
than $6$ months, and identified gold standard app market behaviors. In the
following, we briefly describe the tools we developed, then detail the data
collection effort and the resulting datasets.

%\noindent
{\bf Data collection tools}.
We have developed the \textit{Google Play Crawler} (GPCrawler) tool, to
automatically collect data published by Google Play for apps, users and
reviews.  
Google Play shows only $20$ apps on a user page by default.
GPCrawler overcomes this limitation by using a Firefox add-on and a Python script.
The add-on interacts with Google Play to extend the user page
with a ``scroll down'' button and enable the script to automatically navigate
and collect all the information from the user page.

We have also developed the \textit{Google Play App Downloader} (GPad), a Java tool to automatically download apks of free apps on a PC, using the open-source \textit{Android Market API}~\cite{AndroidMarket}.
%
%and takes as input a list
%of free app ids, a Gmail account and password, and a GSF id.
%
%After collecting the binaries of the apps on the, 
GPad scans each app apk using
VirusTotal~\cite{VirusTotal}, an online malware detector provider,
to find out the number of anti-malware tools (out of $57$:  AVG,
McAfee, Symantec, Kaspersky, Malwarebytes, F-Secure, etc.) that identify the apk as suspicious.
We used $4$ servers (PowerEdge R620, Intel Xeon E-26XX v2 CPUs) to collect our datasets, which we  describe next.

\subsection{Longitudinal App Data.}
\label{sec:data:longitudinal}

%
% for journal
%
%\begin{figure}
%\centering
%\vspace{-5pt}
%\includegraphics[width=0.45\textwidth]{./figures/monitored/apptype.eps}
%\caption{Distribution of app types for the 87,223 fresh app set.  With the
%slight exception of ``Personalization'' and ``Sports'' type spikes, we have
%achieved an almost uniform distribution across all app types, as desirable.
%\bogdan{This figure is ugly and the fonts are too small to read.}}
%\vspace{-15pt}
%\label{fig:fresh:type}
%\end{figure}

%
% for journal
%
%\begin{figure}
%\centering
%\includegraphics[width=2.5in]{./figures/monitored/avgRating87K.eps}
%\vspace{-15pt}
%\caption{Average rating distribution for the 87,223 fresh app set. Most apps
%have more than 3.5 stars, but more than 7000 apps have less than 1 star. \bogdan{This
%plot does not provide enough information to include it now.}}
%\vspace{-25pt}
%\label{fig:fresh:rating}
%\end{figure}

In order to detect app attribute changes that occur early in the lifetime of
apps, we used the ``New Releases'' link to identify apps with a short history
on Google Play.  We approximate the first upload date of an app using the day
of its first review. We have started collecting new releases in July 2014 and
by October 2014 we had a set of $87,223$ apps, whose first upload time was
under $40$ days prior to our first collection time, when they had at most $100$
reviews.
%
%Figure~\ref{fig:fresh:type} shows the distribution of the fresh app types.
%Most apps have at least a $3.5$ rating.  Figure~\ref{fig:fresh:rating} shows the
%average rating distribution of the fresh apps.

We have collected longitudinal data from these $87,223$ apps between October
24, 2014 and May 5, 2015. Specifically, for each app we captured ``snapshots''
of its Google Play metadata, twice a week. An app snapshot consists of values
for all its time varying variables, e.g., the reviews, the rating and install
counts, and the set of requested permissions (see $\S$~\ref{sec:background} for
the complete list).  For each of the $2,850,705$ reviews we have collected from
the $87,223$ apps, we recorded the reviewer's name and id ($2,380,708$ unique
ids), date of review, review title, text, and rating.

%\mizan{For app information table fields are  app id, update/upload date,
%developer name/url/email, apk size, android version, rating value, price. Also
%we have collected reviewers information. We have collected all the apps list of
%those reviewers.}

\subsection{Gold Standard Data.\\}
\label{sec:data:gold}

%
% for journal?
%
%\begin{figure}
%\centering
%\vspace{-15pt}
%\includegraphics[width=2.5in]{figures/clique.15/clique.15.eps}
%\caption{Clique of 15 fraudulent Sybil accounts. Nodes correspond to
%accounts and edges denote the existence of reviews written for the
%same app by two nodes. Edge thickness illustrates the number of apps
%reviewed in common by two accounts.
%\label{fig:clique.15}}
%\vspace{-25pt}
%\end{figure}

\noindent
{\bf Malware apps}.
We used GPad (see $\S$~\ref{sec:data}) to collect the apks of $7,756$ randomly
selected apps from the longitudinal set (see
$\S$~\ref{sec:data:longitudinal}).  Figure~\ref{fig:malware.apps} shows the
distribution of flags raised by VirusTotal, for the $7,756$ apks. None of these
apps had been filtered by Bouncer~\cite{OM12}! From the $523$ apps that were
flagged by at least $3$ tools, we selected those that had at least $10$
reviews, to form our ``malware app'' dataset, for a total of $212$ apps.

\noindent
{\bf Fraudulent apps}.
We used contacts established among Freelancer~\cite{Freelancer}'s search
rank fraud community, to obtain the identities of $15$ Google Play accounts that
were used to write fraudulent reviews. We call these ``seed fraud accounts''.
These accounts were used to review $201$ unique apps. We call these, the ``seed
fraud apps'', and we use them to evaluate FairPlay.
%
%
% for journal
%
%Figure~\ref{fig:clique.15} shows the graph formed by the review
%habits of the 15 seed accounts: nodes are accounts and an edge between two
%accounts encodes the existence of app commonly reviewed by the accounts.

\noindent
{\bf Fraudulent reviews}.
We have collected all the $53,625$ reviews received by the $201$ seed fraud
apps. The $15$ seed fraud accounts were responsible for $1,969$ of these
reviews. We used the $53,625$ reviews to identify $188$ accounts, such
that each account was used to review at least $10$ of the $201$ seed fraud apps
(for a total of $6,488$ reviews). We call these, {\it guilt by association} (GbA)
accounts.
To reduce feature duplication, we have used the $1,969$ fraudulent reviews
written by the $15$ seed accounts and the $6,488$ fraudulent reviews written by the
$188$ GbA accounts for the $201$ seed fraud apps, to extract a {\it balanced} set
of fraudulent reviews.  Specifically, from this set of $8,457$
($=1,969+6,488$) reviews, we have collected $2$ reviews from each of the $203$
($=188+15$) suspicious user accounts. Thus, the gold standard dataset of
fraudulent reviews consists of $406$ reviews.

%The reason for collecting a small number of reviews from each fraudster is to
%reduce feature duplication: many of the features we use to classify a review
%are extracted from the user who wrote the review (see
%Table~\ref{table:review}).

\noindent
{\bf Benign apps}.
We have selected $925$ candidate apps from the longitudinal app set, that have
been developed by Google designated ``top developers''. We have used GPad to
filter out those flagged by VirusTotal. We have manually investigated $601$ of
the remaining apps, and selected a set of $200$ apps that (i) have more than
$10$ reviews and (ii) were developed by reputable media outlets (e.g., NBC,
PBS) or have an associated business model (e.g., fitness trackers).

\noindent
{\bf Genuine reviews}.
We have manually collected a gold standard set of $315$ genuine reviews, as
follows. First, we have collected the reviews written for apps installed on the
Android smartphones of the authors. We then used Google's text and reverse
image search tools to identify and filter those that plagiarized other reviews
or were written from accounts with generic photos. We have then manually
selected reviews that mirror the authors' experience, have at least $150$
characters, and are informative (e.g., provide information about bugs, crash
scenario, version update impact, recent changes).

%, game characters and level difficulty).

\section{FairPlay: Proposed Solution}
\label{sec:fair:play}

%\begin{figure}
%\centering
%\includegraphics[width=0.49\textwidth]{figures/fair.play/feature.table.eps}
%\caption{FairPlay's most important features, organized by the module
%that extracts them (CoReG, RF, IRR and JH modules).
%\label{table:fair:play}}
%\vspace{-25pt}
%\end{figure}

\begin{table}
\begin{center}
\textsf{
\small
\begin{tabular}{l l}
\cmidrule[0.005em](lr{.75em}){1-2}
\textbf{Notation} & \textbf{Definition}\\
\cmidrule[0.005em](lr{.75em}){1-2}
\footnotesize{{\bf CoReG Module}} &\\
$nCliques$ & number of pseudo cliques with $\rho \ge \theta$\\
stats($\rho$) & clique density: max, median, SD\\
stats($cliqueSize$) & pseudo cliques size: max, median, SD \\
$inCliqueSize$ & \% of nodes involved in pseudo cliques\\
\cmidrule[0.005em](lr{.75em}){1-2}
\footnotesize{{\bf RF Module}} &\\
$malW$ & \% of reviews with malware indicators\\
$fraudW$, $goodW$ & \% of reviews with fraud/benign words\\
$FRI$ & fraud review impact on app rating\\
\cmidrule[0.005em](lr{.75em}){1-2}
\footnotesize{{\bf IRR Module}} &\\
stats($spikes$) & days with spikes \& spike amplitude\\
$I_1/Rt_1$, $I_2/Rt_2$ & install to rating ratios\\
$I_1/Rt_1$, $I_2/Rt_2$ & install to review ratios\\
\cmidrule[0.005em](lr{.75em}){1-2}
\footnotesize{{\bf JH Module}} &\\
$permCt$, $dangerCt$ & \# of dangerous and total permissions\\
$rampCt$ & \# of dangerous permission ramps\\
$dangerRamp$ & \# of dangerous permissions added\\
\cmidrule[0.005em](lr{.75em}){1-2}
\end{tabular}
}
\end{center}
\vspace{-15pt}
\caption{FairPlay's most important features, organized by their extracting
module.}
\label{table:fair:play}
\vspace{-20pt}
\end{table}

%We introduce FairPlay, a system to automatically detect malicious
%and fraudulent apps.

\subsection{FairPlay Overview.}

FairPlay organizes the analysis of longitudinal app data into the following 4
modules, illustrated in Figure~\ref{fig:fair:play}. The Co-Review Graph (CoReG)
module identifies apps reviewed in a contiguous time window by groups of users
with significantly overlapping review histories.  The Review Feedback (RF)
module exploits feedback left by genuine reviewers, while the Inter Review
Relation (IRR) module leverages relations between reviews, ratings and install
counts.  The Jekyll-Hyde (JH) module monitors app permissions, with a focus on
dangerous ones, to identify apps that convert from benign to malware. Each
module produces several features that are used to train an app classifier.
FairPlay also uses general features such as the app's average rating, total
number of reviews, ratings and installs, for a total of $28$ features.
Table~\ref{table:fair:play} summarizes the most important features. In the
following, we detail each module and the features it extracts.

%\vspace{-10pt}

\subsection{The Co-Review Graph (CoReG) Module.}

\begin{figure}
\centering
\includegraphics[width=0.47\textwidth]{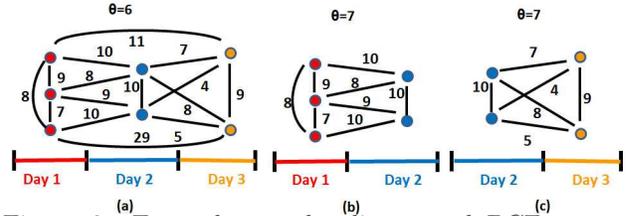}
\vspace{-15pt}
\caption{Example pseudo cliques and PCF output. Nodes are users and edge weights
denote the number of apps reviewed in common by the end users. Review timestamps
have a 1-day granularity. (a) The entire co-review graph, detected as
pseudo clique by PCF when $\theta$ is 6.  When $\theta$ is 7, PCF detects the
subgraphs of (b) the first two days and (c) the last two days.
%
%When $\theta$=8, PCF detects only the clique formed by the
%first day reviews (the red nodes).
\label{fig:ncf}}
\vspace{-20pt}
\end{figure}

Let the {\it co-review graph} of an app, see Figure~\ref{fig:ncf}, be a graph
where nodes correspond to users who reviewed the app, and undirected edges have
a weight that indicates the number of apps reviewed in common by the edge's
endpoint users. We seek to identify cliques in the co-review graph.
Figure~\ref{fig:clique} shows the co-review clique of one of the seed fraud
apps (see $\S$~\ref{sec:data:gold}).

To address the problem's NP-hardness, we exploit two observations.  First,
fraudsters hired to review an app are likely to post those reviews within
relatively short time intervals (e.g., days).  Second, perfect cliques are not
necessary. Instead, we relax this requirement to identify ``pseudo cliques'',
or groups of highly but not necessarily completely connected nodes.
Specifically, we use the weighted density definition of Uno~\cite{U07}: given a
co-review graph, its weighted density $\rho = \frac{\sum_{e \in E}
w(e)}{{n\choose 2}}$, where $E$ denotes the graph's edges and $n$ its number of
nodes (reviews). We are interested then in subgraphs of the co-review graph
whose weighted density exceeds a threshold value $\theta$.

%$\rho$ is thus the average weight of the graph's edges, normalized by the
%total number of edges of a perfect clique of size $n$.

\begin{figure}[h]
%\vspace{-10pt}
\renewcommand{\baselinestretch}{0.9}
\begin{minipage}{0.4\textwidth}
\begin{algorithm}[H]
\begin{tabbing}
XXX\=X\=X\=X\=X\=X\= \kill

\small{{\bf Input:} $days$, an array of daily\ reviews, and}\\
\small{\qquad\qquad $\theta$, the weighted threshold density}\\
\small{{\bf Output:} $allCliques$, set of all detected pseudo cliques}\\

1.\ \small{\mbox{\bf{for}}\ d :=0\;\ d $<$ days.size();\ d++}\\
2.\> \small{Graph\ PC := new\ Graph();}\\
3.\> \small{bestNearClique(PC, days[d]);}\\
4.\> \small{c := 1; n := PC.size();}\\
5.\> \small{\mbox{\bf{for}}\ nd := d+1;\ d $<$ days.size()\ \&\ c = 1;\ d++}\\
6.\>\> \small{bestNearClique(PC, days[nd]);}\\
7.\>\> \small{c := (PC.size() $>$ n);\ \mbox{\bf{endfor}}}\\
8.\> \small{\mbox{\bf{if}}\ (PC.size() $>$ 2)}\\
9.\>\> \small{allCliques := allCliques.add(PC); \mbox{\bf{fi}}\ \mbox{\bf{endfor}}}\\
10.\ \small{return}\\

11.\ \small{\mbox{\bf{function}}\ \underline{bestNearClique}(Graph\ PC, Set\ revs)}\\
12.\>\small{\mbox{\bf{if}}\ (PC.size() = 0)}\\
13.\>\>\small{\mbox{\bf{for}}\ root := 0;\ root $<$ revs.size();\ root++}\\
14.\>\>\>\small{Graph\ candClique := new\ Graph ();}\\
15.\>\>\>\small{candClique.addNode (root.getUser());}\\
16.\>\>\>\small{\mbox{\bf{do}}\ candNode := getMaxDensityGain(revs);}\\
17.\>\>\>\>\small{\mbox{\bf{if}}\ (density(candClique $\cup$ \{candNode\}) $\ge$ $\theta$))}\\
18.\>\>\>\>\>\small{candClique.addNode(candNode); \mbox{\bf{fi}}}\\
19.\>\>\>\small{\mbox{\bf{while}}\ (candNode != null);}\\
20.\>\>\>\small{\mbox{\bf{if}}\ (candClique.density() $>$ maxRho)}\\
21.\>\>\>\>\>\small{maxRho := candClique.density();}\\
22.\>\>\>\>\>\small{PC := candClique; \mbox{\bf{fi}}\ \mbox{\bf{endfor}}}\\
23.\>\small{\mbox{\bf{else\ if}}\ (PC.size() $>$ 0)}\\
24.\>\>\>\small{\mbox{\bf{do}}\ candNode := getMaxDensityGain(revs);}\\
25.\>\>\>\>\small{\mbox{\bf{if}}\ (density(candClique $\cup$ candNode) $\ge$ $\theta$))}\\
26.\>\>\>\>\>\small{PC.addNode(candNode); \mbox{\bf{fi}}}\\
27.\>\>\>\small{\mbox{\bf{while}}\ (candNode != null);}\\
28.\ \small{return}
\vspace{-10pt}
\end{tabbing}
\caption{PCF algorithm pseudo-code.}
\label{alg:cf}
\end{algorithm}
\end{minipage}
\normalsize
\vspace{-20pt}
\end{figure}

We present the Pseudo Clique Finder (PCF) algorithm (see
Algorithm~\ref{alg:cf}), that takes as input the set of the reviews of an app,
organized by days, and a threshold value $\theta$. PCF outputs a set of
identified pseudo cliques with $\rho \ge \theta$, that were formed during
contiguous time frames. In Section~\ref{sec:evaluation:app} we discuss the
choice of $\theta$.

For each day when the app has received a review (line $1$), PCF finds the day's
most promising pseudo clique (lines $3$ and $12-22$): start with each review,
then greedily add other reviews to a candidate pseudo clique; keep the pseudo
clique (of the day) with the highest density. With that ``work-in-progress''
pseudo clique, move on to the next day (line $5$): greedily add other reviews
while the weighted density of the new pseudo clique equals or exceeds $\theta$
(lines $6$ and $23-27$).  When no new nodes have been added to the
work-in-progress pseudo clique (line $8$), we add the pseudo clique to the
output (line $9$), then move to the next day (line $1$). The greedy choice
($getMaxDensityGain$, not depicted in Algorithm~\ref{alg:cf}) picks the review
not yet in the work-in-progress pseudo clique, whose writer has written the
most apps in common with reviewers already in the pseudo clique.
Figure~\ref{fig:ncf} illustrates the output of PCF for several $\theta$ values.

If $d$ is the number of days over which $A$ has received reviews and $r$ is the
maximum number of reviews received in a day, PCF's complexity is
$O(dr^2(r+d))$.

\noindent
{\bf CoReG features}.
CoReG extracts the following features from the output of PCF (see
Table~\ref{table:fair:play}) (i) the number of cliques whose density equals or
exceeds $\theta$, (ii) the maximum, median and standard deviation of the
densities of identified pseudo cliques, (iii) the maximum, median and standard
deviation of the node count of identified pseudo cliques, normalized by $n$
(the app's review count), and (iv) the total number of nodes of the co-review
graph that belong to at least one pseudo clique, normalized by $n$.

\begin{figure*}
\captionsetup[subfigure]{aboveskip=0.1in, belowskip=-0.05in}
\begin{subfigure}[b]{2.05in}
\includegraphics[width=1.85in]{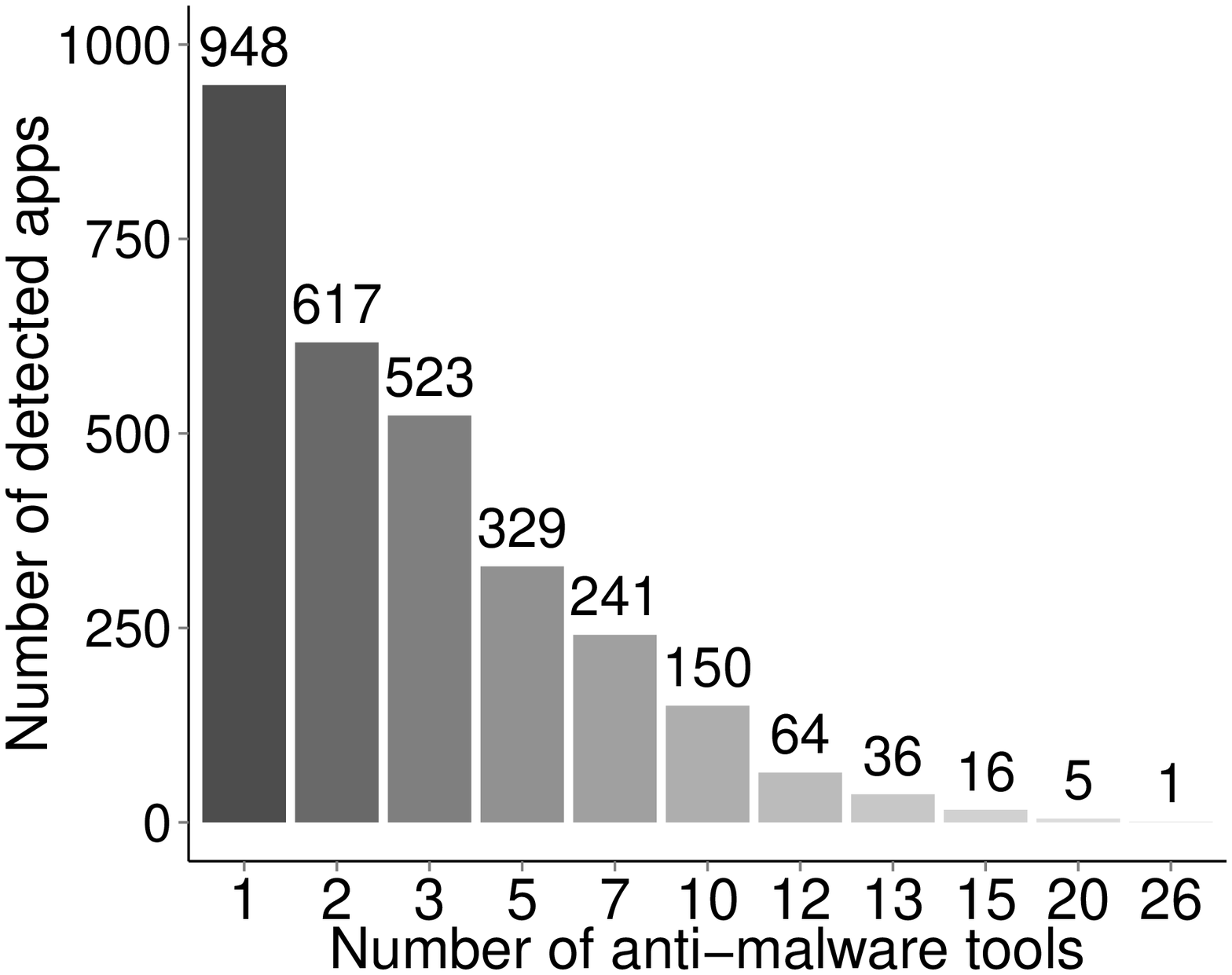}
\caption{}
\label{fig:malware.apps}
\end{subfigure}
~
\begin{subfigure}[b]{2.50in}
\includegraphics[height=1.59in]{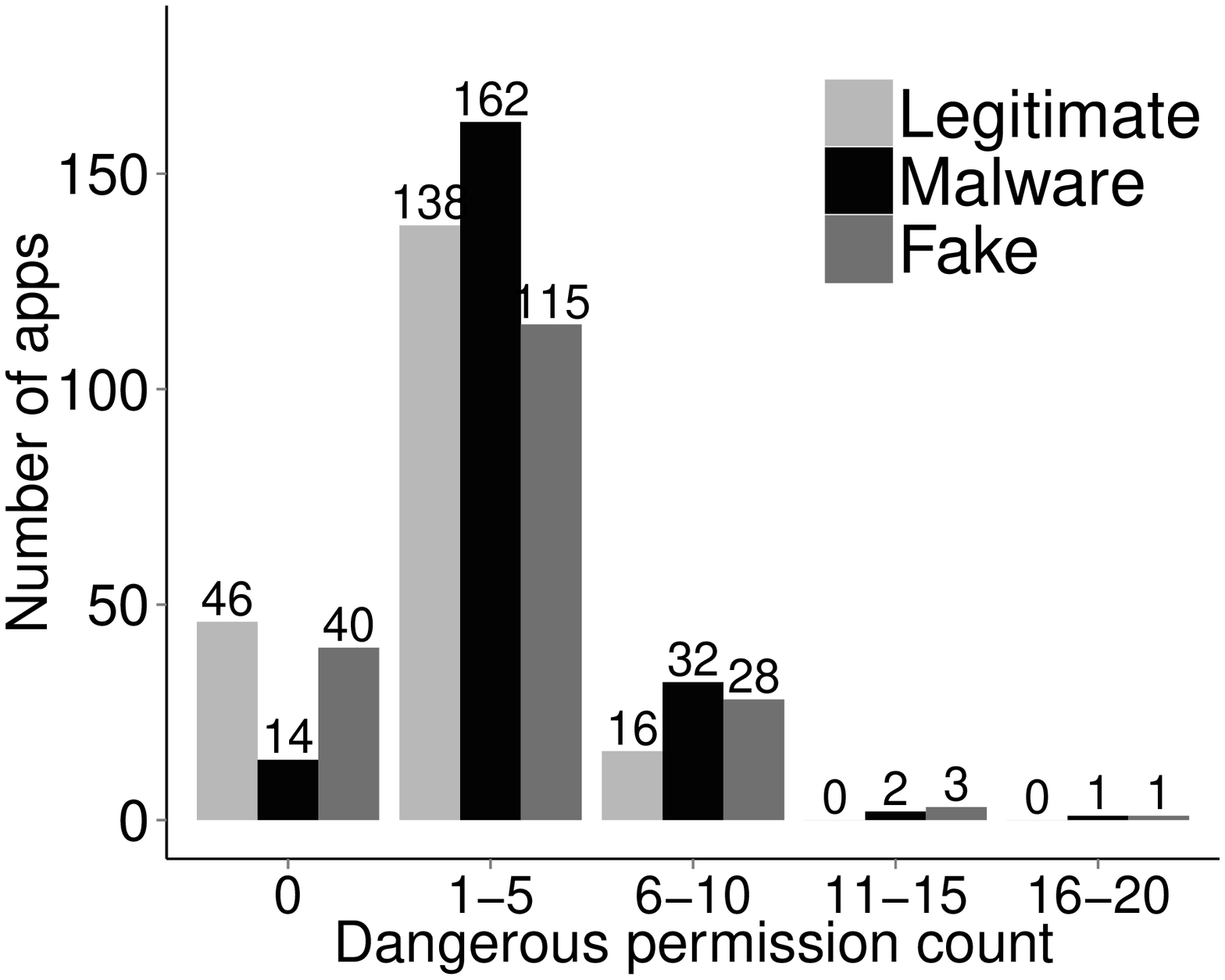}
\caption{}
\label{fig:permissions:dangerous}
\end{subfigure}
~
\begin{subfigure}[b]{2.05in}
\includegraphics[width=2.05in]{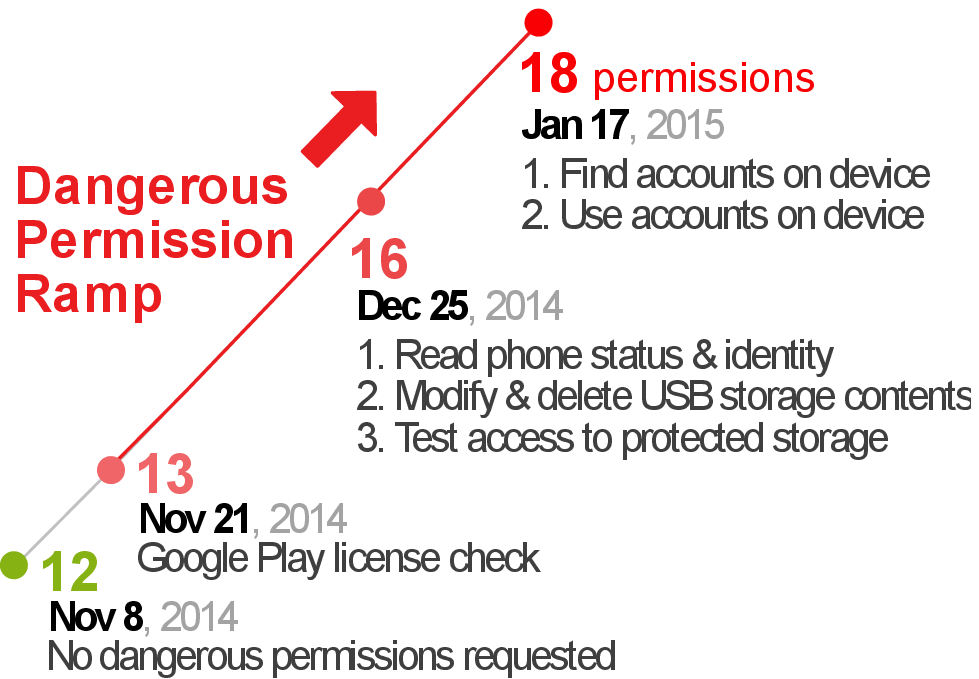}
\caption{}
\label{fig:permissions:ramp}
\end{subfigure}
\caption{(a) Apks detected as suspicious ($y$ axis) by multiple
anti-virus tools ($x$ axis), through VirusTotal~\cite{VirusTotal},
from a set of $7,756$ downloaded apks.
(b) Distribution of the number of ``dangerous'' permissions requested by
malware, fraudulent and benign apps.
(c) Dangerous permission ramp during version updates for a sample app
``com.battery.plusfree''. Originally the app requested no dangerous
permissions.}
\vspace{-15pt}
\end{figure*}

%\vspace{-10pt}

\subsection{Reviewer Feedback (RF) Module.}

%\begin{table}
%\begin{center}
%\textsf{
%\small
%\begin{tabular}{l l}
%\toprule
%\textbf{Notation} & \textbf{Definition}\\
%\midrule
%$\rho_R$ & The rating of $R$\\
%$L(R)$ & The length of $R$\\
%$pos(R)$ & Percentage of positive statements in $R$\\
%$neg(R)$ & Percentage of negative statements in $R$\\
%\midrule
%$nr(U)$ & The number of reviews written by $U$\\
%$\pi(\rho_R)$ & Percentile of $\rho_R$ among all reviews of $U$\\
%$Exp_U(A)$ & The expertise of $U$ for app $A$\\
%$B_U(A)$ & The bias of $U$ for $A$\\
%$Paid(U)$ & The money spent by $U$ to buy apps\\
%$Rated(U)$ & Number of apps rated by $U$\\
%$plusOne(U)$ & Number of apps +1'd by $U$\\
%$n.flwrs(U)$ & Number of followers of $U$ in Google+\\
%\bottomrule
%\end{tabular}
%}
%\end{center}
%\vspace{-15pt}
%\caption{Features used to classify review $R$ written by user $U$ for
%app $A$.}
%\label{table:review}
%\vspace{-20pt}
%\end{table}

Reviews written by genuine users of malware and fraudulent apps may describe
negative experiences. The RF module exploits this observation through a two
step approach: (i) detect and filter out fraudulent reviews, then (ii) identify
malware and fraud indicative feedback from the remaining reviews.

\noindent
{\bf Step RF.1: Fraudulent review filter}.
We posit that users that have higher expertise on apps they review, have
written fewer reviews for apps developed by the same developer, have reviewed
more paid apps, are more likely to be genuine. We exploit this conjecture to
use supervised learning algorithms trained on the following features, defined
for a review $R$ written by user $U$ for an app $A$:

\noindent
$\bullet$
{\it Reviewer based features}.
The \textit{expertise} of $U$ for app $A$, defined as the number of reviews $U$
wrote for apps that are ``similar'' to $A$, as listed by Google Play (see
$\S$~\ref{sec:background}). The \textit{bias} of $U$ towards $A$: the number of
reviews written by $U$ for other apps developed by $A$'s developer. In
addition, we extract the total money paid by $U$ on apps it has reviewed, the
number of apps that $U$ has liked, and the number of Google+ followers of $U$.

\noindent
$\bullet$
{\it Text based features}.
We used the NLTK library~\cite{NLTK} and the Naive Bayes classifier, trained on
two datasets: (i) $1,041$ sentences extracted from randomly selected $350$
positive and $410$ negative Google Play reviews, and (ii) $10,663$ sentences
extracted from $700$ positive and $700$ negative IMDB movie
reviews~\cite{PangLee}.  $10$-fold cross validation of the Naive Bayes
classifier over these datasets reveals a FNR of $16.1$\% and FPR of $19.65$\%.
%
% for an overall accuracy of $81.74$\%.
%We ran a binomial test~\cite{mcdonald:statistics}
%for a given accuracy of p=0.817 over $N$=1,041 cases using the binomial
%distribution $bionmial(p,N)$ to assess the 95\% confidence interval for our
%result.  The deviation of the binomial distribution is 0.011. Thus, we are 95\%
%confident that the true performance of the classifier is in the interval
%(79.55, 83.85).
%
We used the trained Naive Bayes classifier to determine the statements of $R$
that encode positive and negative sentiments. We then extracted the following
features: (i) the percentage of statements in $R$ that encode positive and
negative sentiments respectively, and (ii) the rating of $R$ and its percentile
among the reviews written by $U$.

%In Section~\ref{sec:evaluation} we evaluate the review classification accuracy
%of these features, on the gold standard datasets of fraudulent and genuine
%reviews introduced in Section~\ref{sec:data:gold}.

\noindent
{\bf Step RF.2: Reviewer feedback extraction}.
We conjecture that (i) since no app is perfect, a ``balanced'' review that
contains both app positive and negative sentiments is more likely to be
genuine, and (ii) there should exist a relation between the review's dominating
sentiment and its rating. Thus, after filtering out fraudulent reviews, we
extract feedback from the remaining reviews. For this, we have used NLTK to
extract $5,106$ verbs, $7,260$ nouns and $13,128$ adjectives from the $97,071$
reviews we collected from the $613$ gold standard apps (see
$\S$~\ref{sec:data:gold}).
%
%we removed non ascii characters
%and stop words, then applied lemmatization and discarded words that appear at
%most once. We ignored stemming due to poor results.
%
We used these words to manually identify lists of words indicative of malware,
fraudulent and benign behaviors. Our malware indicator word list contains $31$
words (e.g., risk, hack, corrupt, spam, malware, fake, fraud, blacklist, ads).
The fraud indicator word list contains $112$ words (e.g., cheat, hideous,
complain, wasted, crash) and the benign indicator word list contains $105$ words.

\noindent
{\bf RF features}.
We extract 3 features (see Table~\ref{table:fair:play}), denoting the
percentage of genuine reviews that contain malware, fraud, and benign indicator
words respectively. We also extract the {\it impact} of detected
fraudulent reviews on the overall rating of the app: the absolute difference
between the app's average rating and its average rating when ignoring all the
fraudulent reviews.

\subsection{Inter-Review Relation (IRR) Module.}

This module leverages temporal relations between reviews, as well as
relations between the review, rating and install counts of apps, to identify
suspicious behaviors.

%
% for journal
%
%\begin{figure}
%\centering
%\includegraphics[width=2.9in]{graphs/reviews/spike3Timeline.eps}
%\caption{Timelines of positive reviews for 2 apps from the fraudulent app
%dataset.}
%\label{fig:review:spikes}
%\end{figure}

%
% for journal
%
%\noindent
%{\bf Step R.2: Review frequency}.
%In order to compensate for a negative review, an attacker needs to post a
%significant number of positive reviews:
%
%\vspace{-5pt}
%
%\begin{claim}
%Let $\rho_A$ denote the rating of an app $A$ just before receiving a 1 star
%review.  In order to compensate for the 1 star review, an attacker needs to
%post at least $\frac{\rho_A-1}{5-\rho_A}$ positive reviews.
%\end{claim}
%
%\vspace{-10pt}
%
%\begin{proof}
%Let $\sigma$ be the sum of all the $k$ reviews received by $a$ before time $T$.
%Then, $\rho_A = \frac{\sigma}{k}$. Let $q_r$ be the number of fraudulent
%reviews received by $A$. To compensate for the 1 star review posted at time $T$, $q_r$
%is minimized when all those reviews are 5 star. We then have that:
%$\rho_A = \frac{\sigma}{k} = \frac{\sigma+1+5q_r}{k+1+q_r}$.  The numerator of
%the last fraction denotes the sum of all the ratings received by $A$ after time
%$T$ and the denominator is the total number of reviews.  Rewriting the last
%equality, we obtain that $q_r = \frac{\sigma - k}{5k - \sigma} = \frac{\rho_a -
%1}{5 - \rho_a}$.  The last equality follows by dividing both the numerator and
%denominator by $k$.
%\end{proof}
%
%For instance, a 4 star app needs to receive 3, 5-star reviews to compensate for
%a single 1 star review.

\noindent
{\bf Temporal relations}.
We detect outliers in the number of daily reviews received by an app.
%Figure~\ref{fig:review:spikes} shows the timelines and suspicious spikes of
%positive reviews for 2 apps from the fraudulent app dataset (see
%Section~\ref{sec:data:gold}).
We identify days with spikes of positive reviews
as those whose number of positive reviews exceeds the upper outer fence of the
box-and-whisker plot built over the app's numbers of daily positive reviews.

\begin{figure}
\centering
\vspace{-15pt}
\includegraphics[width=0.45\textwidth]{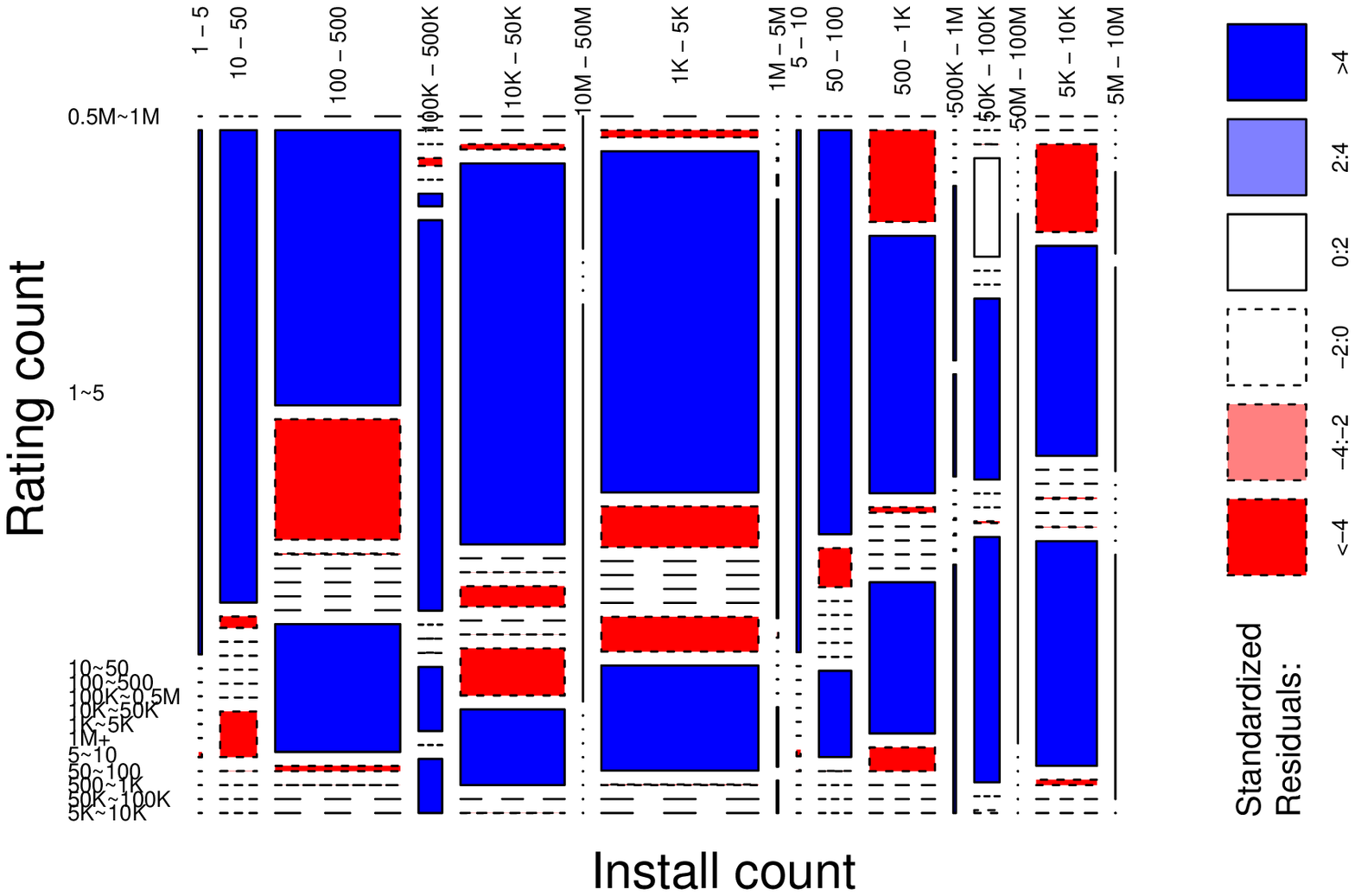}
\vspace{-15pt}
\caption{Mosaic plot of install vs. rating count relations of the $87$K apps.
Larger rectangles signify that more apps have the corresponding rating and
install count range; dotted lines mean no apps in a certain install/rating
category. The standardized residuals identify the cells that contribute the
most to the $\chi^2$ test. The most significant rating:install ratio is 1:100.
%
%Ratios as low as 1 to 10
%also exist but are less significant.
}
\vspace{-20pt}
\label{fig:fresh:mosaic}
\end{figure}

%
% for journal
%
%\begin{figure}
%\centering
%\includegraphics[width=0.5\textwidth]{./figures/monitored/mosaic.install.rating.eps}
%\caption{Mosaic plot showing relationships between the install count and the average
%rating of the fresh apps. It shows that as apps receive more installs, they also tend
%to have higher average rating. We believe this is because on average, apps tend
%to be more popular if they are good (higher rating). Also, a higher rating also
%leads to improved install count.}
%\label{fig:fresh:mosaic:rating}
%\end{figure}

\noindent
{\bf Reviews, ratings and install counts}.
We used the Pearson's $\chi^2$ test to investigate relationships between the
install and rating counts of the $87$K new apps, at the end of the collection
interval.  We grouped the rating count in buckets of the same size as Google
Play's install count buckets.  Figure~\ref{fig:fresh:mosaic} shows the mosaic
plot of the relationships between rating and install counts. $p$=$0.0008924$,
thus we conclude dependence between the rating and install counts.  We leverage
this result to conjecture that adversaries that post fraudulent ratings and
reviews, or create fake app install events, may break a natural balance
between their counts.

\noindent
{\bf IRR features}.
We extract temporal features (see Table~\ref{table:fair:play}): the number of
days with detected spikes and the maximum amplitude of a spike. We also extract
(i) the ratio of installs to ratings as two features,
$I_1/Rt_1$ and $I_2/Rt_2$ and (ii) the ratio of installs to reviews, as
$I_1/Rv_1$ and $I_2/Rv_2$. $(I_1, I_2]$ denotes the install count interval of
an app, $(Rt_1, Rt_2]$ its rating interval and $(Rv_1, Rv_2]$ its (genuine)
review interval.

%Googl Play restricts the number of displayed reviews, to 4000 per app. While
%our fresh apps have fewer reviews, we note that our approach, of daily
%monitoring the apps, solves this problem even for apps with more reviews: each
%day we only collect the reviews posted from the previous day. We now have more
%than 300,000 reviews (after roughly 1 month of fresh app monitoring).  For
%each
%review we record the reviewer name, reviewer URL, date of review, review title,
%review text, and rating value.

%8. Information about the developer. Needs discussion. For instance, if the
%developer has url, run that url through virus total. Also, how many apps from
%our set has the developer developed?
%\ratul{To me, this is optional. Will be the last thing to check if time permits.}

%\vspace{-10pt}

\subsection{Jekyll-Hyde App Detection (JH) Module.}

%
% for journal
%
%\begin{figure}
%\centering
%\includegraphics[width=2.5in]{graphs/permissions/permissions.all.eps}
%\caption{Distribution of total number of permissions requested by malware,
%fraudulent and legitimate apps.}
%\label{fig:permissions:all}
%\end{figure}

%\begin{figure}
%\begin{subfigure}[b]{1.50in}
%\includegraphics[width=1.50in]{graphs/permissions/permissions.dangerous.eps}
%\caption{}
%\label{fig:permissions:dangerous}
%\end{subfigure}
%~
%\begin{subfigure}[b]{1.80in}
%\includegraphics[width=1.80in]{graphs/permissions/ramp.eps}
%\caption{}
%\label{fig:permissions:ramp}
%\end{subfigure}
%\caption{(a) Distribution of the number of ``dangerous'' permissions requested by
%malware, fraudulent and legitimate apps.
%(b) Dangerous permission ramp during version updates for a sample app
%"com.battery.plusfree". Originally the app requested no dangerous
%permissions.}
%\end{figure}

%\begin{figure}
%\centering
%\includegraphics[width=2.9in]{graphs/permissions/permissions.dangerous.eps}
%\caption{Distribution of the number of ``dangerous'' permissions requested by
%malware, fraudulent and legitimate apps. The top most popular dangerous
%permissions among our datasets of 601 malware, fraudulent and legitimate apps
%are ``modify or delete the contents of the USB storage'' (401 apps), ``read
%phone status and identity'', ``find accounts on the device'', and ``access
%precise location''.}
%\label{fig:permissions:dangerous}
%\vspace{-15pt}
%\end{figure}

Android's API level 22 labels $47$ permissions as ``dangerous''.
Figure~\ref{fig:permissions:dangerous} compares the distributions of the number
of dangerous permissions requested by the gold standard malware, fraudulent and
benign apps.  The  most popular dangerous permissions among these apps are
``modify or delete the contents of the USB storage'', ``read phone
status and identity'', ``find accounts on the device'', and ``access precise
location''. Most benign apps request at most $5$ such permissions; some
malware and fraudulent apps request more than $10$.

%\begin{figure}
%\centering
%\includegraphics[width=2.9in]{graphs/permissions/ramp.eps}
%\caption{Dangerous permission ramp during version updates for a sample app
%"com.battery.plusfree". Originally the app requested no dangerous
%permissions.
%\label{fig:permissions:ramp}}
%\end{figure}

%After the recent Google Play policy change~\cite{PermissionSimplify}, Google
%Play now groups app permissions into groups of related permissions. Apps can
%request a group of permissions and gain implicit access to dangerous
%permissions.
%
Upon manual inspection of several apps, we identified a new type of malicious
intent possibly perpetrated by deceptive app developers: apps that seek to
attract users with minimal permissions, but later request dangerous
permissions. The user may be unwilling to uninstall the app ``just'' to reject
a few new permissions. We call these {\it Jekyll-Hyde apps}. Figure
~\ref{fig:permissions:ramp} shows the dangerous permissions added during
different version updates of one gold standard malware app.

\noindent
{\bf JH features}.
We extract the following features (see Table~\ref{table:fair:play}), (i) the
total number of permissions requested by the app, (ii) its number of dangerous
permissions, (iii) the app's number of dangerous permission ramps, and (iv) its
total number of dangerous permissions added over all the ramps.

%\begin{claim}
%Assuming a Poisson arrival process for positive reviews, $X$, the probability
%that the number of positive reviews received in a $\delta t$ interval exceeds
%the average number of positive reviews $\alpha$ by a factor of $x$ is $Pr[X \ge x \alpha] =
%\sum_{i=\lceil \frac{nx \delta t}{\Delta T} \rceil}^{\infty} \frac{(\frac{n \delta t}{\Delta T})^i e^{-\frac{n \delta t}{\Delta T}}}{i!}$.
%\end{claim}
%
%\begin{proof}
%
%The average number of positive reviews received in an interval $\delta t$ is
%$\alpha = \lambda \delta t$. $\lambda = \frac{n}{\Delta T}$. Then,
%\[
%Pr[X \ge x \alpha] = Pr[X \ge \lceil \frac{nx \delta t}{\Delta T} \rceil] =
%\]
%\[
%\sum_{i = x \rceil \frac{nx \delta t}{\Delta T} \rceil}^{\infty} Pr[X = i] =
%\sum_{i = x \rceil \frac{nx \delta t}{\Delta T} \rceil}^{\infty} \frac{(\frac{n \delta t}{\Delta T})^i e^{-\frac{n \delta t}{\Delta T}}}{i!}
%\]
%
%\end{proof}

\section{Evaluation}
\label{sec:evaluation}

\subsection{Experiment Setup.}
\label{sec:evaluation:setup}

We have implemented FairPlay using Python to extract data from parsed pages and
compute the features, and the R tool to classify reviews and apps.  We have set
the threshold density value $\theta$ to 3, to detect even the smaller pseudo
cliques.

We have used the Weka data mining suite~\cite{Weka} to perform the experiments,
with default settings. 
We experimented with multiple supervised learning
algorithms.
Due to space constraints, we report results for the best performers: MultiLayer Perceptron
(MLP)~\cite{MLP}, Decision Trees (DT) (C4.5) and Random Forest
(RF)~\cite{RF2001}, using $10$-fold cross-validation~\cite{K95}.
%
%to asses how the results of the statistical analysis will
%generalize to an independent data set.
%For the
%backpropagation algorithm of the MLP classifier, we set the learning rate to
%0.3 and the momentum rate to 0.2. We used MySQL to store collected
%data and features.
We use the term ``positive'' to denote a fraudulent review,
fraudulent or malware app; FPR means \textit{false positive rate}.
Similarly, ``negative'' denotes a genuine review or
benign app; FNR means \textit{false negative rate}. 

%In the following, we use the term ``positive'' to denote a fraudulent review,
%fraudulent or malware app, and ``negative'' to denote a genuine review or
%benign app. The false positive rate (FPR) denotes the probability of falsely
%classifying an instance as positive, while the false negative rate (FNR)
%denotes the chance of falsely classifying an instance as negative. A classifier's
%accuracy is the ratio of true positives and negatives to the entire population.

%We use the Receiver Operating Characteristic (ROC) curve to visually display
%the trade-off between the false positive rate (FPR) and false negative rate
%(FNR).  TPR is the true positive rate. The Equal Error Rate (EER) is the rate
%at which both positive and negative errors are equal. A lower EER denotes a
%more accurate solution.

\subsection{Review Classification.}
\label{sec:evaluation:review}

\begin{table}
\setlength{\tabcolsep}{.16667em}
\centering
\textsf{
\begin{tabular}{l | r r | r}
\toprule
\textbf{Strategy} & \textbf{FPR}\% & \textbf{FNR}\% & \textbf{Accuracy}\%\\
\midrule
%Bagging & 94.29 & 3.4 & 5.71 & 95.57\\
DT {(\small Decision Tree)}  & 2.46 & 6.03 & 95.98\\
%{\color{red}RF.1/MLP} & {\color{red}$1.47$} & {\color{red}$6.67$} & {\color{red}$96.26$}\\
MLP {(\small Multi-layer Perceptron)}  & \textbf{1.47} & 6.67 & 96.26\\
RF {(\small Random Forest)}  & 2.46 & 5.40 & 96.26\\
\bottomrule
\end{tabular}
}
\vspace{-5pt}
\caption{Review classification results ($10$-fold cross-validation), of gold standard fraudulent (positive) and genuine (negative) reviews. 
MLP achieves the lowest false positive rate (FPR) of $1.47$\%.}
\vspace{-20pt}
\label{table:review:classification}
\end{table}

%
% for journal
%
%\begin{figure}
%\centering
%\includegraphics[width=2.9in]{graphs/reviews/ROC.reviews.eps}
%\caption{ROC plot of 3 classifiers: Decision Tree, Random Forest and
%Bagging for review classification. RF and MLP are tied for best accuracy, of $96.26$\%.
%The EER of MLP is as low as 0.036.
%\label{fig:reviews:roc}}
%\end{figure}

To evaluate the accuracy of FairPlay's fraudulent review detection component
(RF module), we used the gold standard datasets of fraudulent and genuine
reviews of $\S$~\ref{sec:data:gold}. We used GPCrawler to collect the data of
the writers of these reviews, including the $203$ reviewers of the $406$
fraudulent reviews ($21,972$ reviews for $2,284$ apps) and the $315$ reviewers
of the genuine reviews ($9,468$ reviews for $7,116$ apps).
%
%We observe that the users who post genuine reviews write fewer reviews in total
%than those who post fraudulent reviews; however, overall, those users review
%more apps apps in total.
%We have also collected information about each of these collected apps, e.g.,
%the identifiers of the app developer. 
%
Table~\ref{table:review:classification} shows the results of the $10$-fold
cross validation of algorithms classifying reviews as genuine or fraudulent.
To minimize wrongful accusations, we seek to minimize the FPR~\cite{CNWWF11}.
MLP simultaneously achieves the highest accuracy of $96.26\%$ and the lowest FPR of $1.47$\%  (at $6.67$\% FNR).
%(EER is $3.6$\%, AUC=$0.98$). 
Thus, in the following experiments, we use MLP to filter
out fraudulent reviews in the RF.1 step.

%
% for journal: app classification ROC plot
%
%\begin{figure}
%	\centering
%	\includegraphics[width=2.9in]{graphs/apps/ROC.apps.fake.eps}
%	\caption{ROC plot of 3 classifiers: Decision Tree, MLP and
%		Bagging for app classification (legitimate vs fake). DT has the highest accuracy, of 98.99\%. 
%		The EER of MLP is as low as 0.01.
%		\label{fig:apps:roc}}
%\end{figure}

%The top 3 most
%impactful features for MLP are (i) the maximum deviation of the node count of
%identified pseudo cliques, normalized by n, (ii) the total number of nodes that
%belong to at least one identified pseudo clique, normalized by n and (iii) the
%number of spikes. \bogdan{Are these features for the review or for the app
%classification?  Do we use cliques to classify reviews?}

\subsection{App Classification\\}
\label{sec:evaluation:app}

\begin{table}[t]
\centering
\textsf{
\begin{tabular}{l | r r | r}
\toprule
\textbf{Strategy} & \textbf{FPR}\% & \textbf{FNR}\% & \textbf{Accuracy}\%\\
\midrule
%Bagging & $97.48$ & $1.51$ & $2.52$ & $97.99$\\
%{\color{red}FairPlay/RF} & {\color{red}$1.01$} & {\color{red}$3.52$} & {\color{red}$97.74$}\\
FairPlay/DT & 3.01 & 3.01 & 96.98\\
FairPlay/MLP & 1.51 & 3.01 & 97.74\\
FairPlay/RF & \textbf{1.01} & 3.52 & 97.74\\
\bottomrule
\end{tabular}
}
\vspace{-5pt}
\caption{FairPlay classification results ($10$-fold cross validation) of gold standard
fraudulent (positive) and benign apps. 
RF has lowest FPR, thus desirable~\cite{CNWWF11}.}
\vspace{-5pt}
\label{table:fp:fraudulent}
\end{table}

%\begin{figure}
%\centering
%\includegraphics[width=1.9in]{./figures/clique.15/clique.eps}
%\vspace{-5pt}
%\caption{Clique detected by PCF for ``Tiempo - Clima gratis'', one of the 201
%seed fraud apps (see $\S$~\ref{sec:data:gold}). Edge width is proportional to
%its weight. The minimum edge weight in this 37 node clique is 115 (number of
%apps reviewed in common by two users) and the maximum weight is 164.}
%\label{fig:clique}
%\vspace{-15pt}
%\end{figure}

\begin{figure*}
\captionsetup[subfigure]{aboveskip=0.05in, belowskip=-0.08in}
\begin{subfigure}[b]{1.75in}
\includegraphics[height=1.65in]{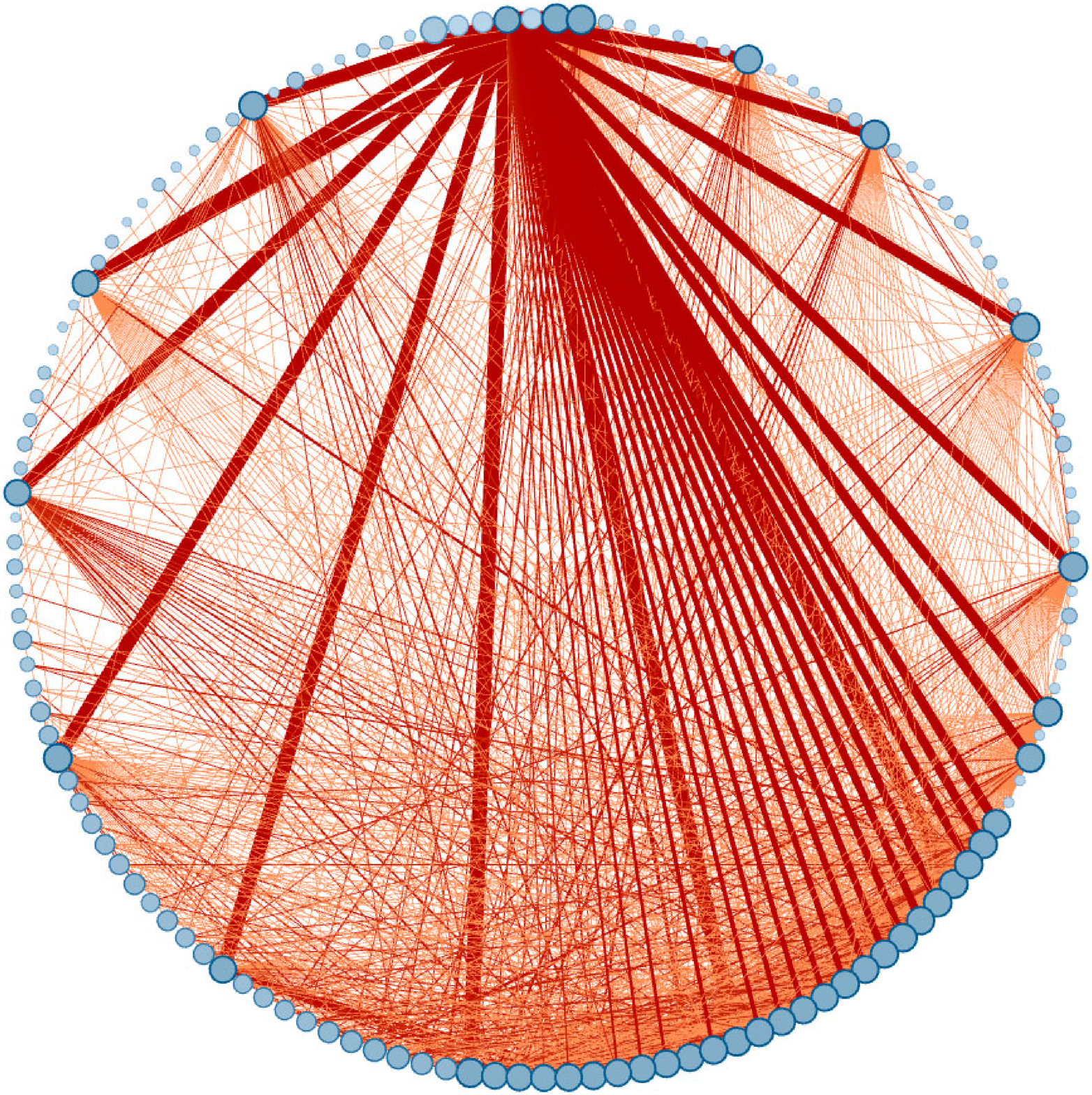}
\caption{}
\label{fig:clique}
\end{subfigure}
~
\captionsetup[subfigure]{aboveskip=0.05in, belowskip=-0.08in}
\begin{subfigure}[b]{2.35in}
\includegraphics[width=2.1in]{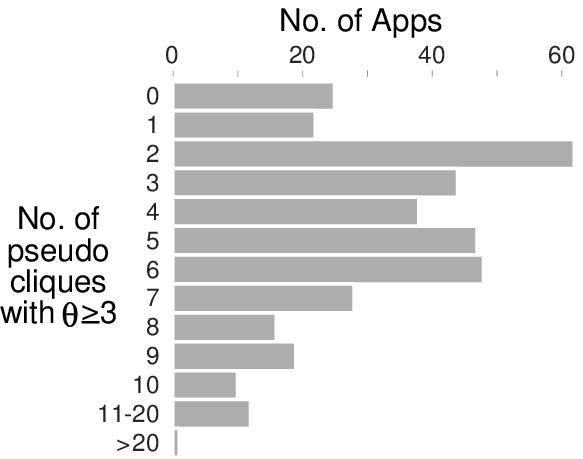}
\caption{}
\label{fig:clique:count}
\end{subfigure}
~
\captionsetup[subfigure]{aboveskip=0.05in, belowskip=-0.08in}
\begin{subfigure}[b]{2.35in}
\includegraphics[width=2.35in]{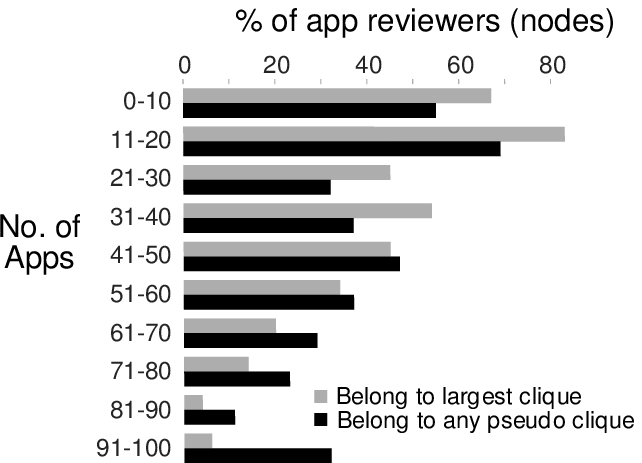}
\caption{}
\label{fig:clique:members}
\end{subfigure}
\caption{
(a) Clique flagged by PCF for ``Tiempo - Clima gratis'', one of the $201$ seed
fraud apps (see $\S$~\ref{sec:data:gold}), involving $37$ reviewers (names
hidden for privacy); edge weights proportional to numbers of apps reviewed in
common (ranging from $115$ to $164$ apps).
(b \& c) Statistics over the $372$ fraudulent apps out of $1,600$ investigated:
(b) Distribution of per app number of discovered pseudo cliques. 
$93.3$\% of the $372$ apps have at least $1$ pseudo clique of $\theta \ge 3$
(c) Distribution of percentage of app reviewers (nodes) that belong to the
largest pseudo clique and to any clique. $8$\% of the $372$ apps have more than
$90$\% of their reviewers involved in a clique!
}
\vspace{-15pt}
\end{figure*}

To evaluate FairPlay, we have collected all the $97,071$ reviews of the $613$
gold standard malware, fraudulent and benign apps, written by $75,949$ users,
as well as the $890,139$ apps rated or played by these users.

\noindent
{\bf Fraud Detection Accuracy}.
Table~\ref{table:fp:fraudulent} shows $10$-fold cross validation results of
FairPlay on the gold standard fraudulent and benign apps (see
$\S$~\ref{sec:data:gold}).  
All classifiers achieve accuracies of around $97$\%.
Random Forest is the best, having the highest accuracy of $97.74\%$ and the lowest FPR of $1.01$\%.
%with an EER of $2.5$\% and the area under the ROC curve (AUC) of $0.993$. 

Figure~\ref{fig:clique} shows the co-review subgraph for one of the seed fraud
apps identified by FairPlay's PCF. We observe that the app's reviewers form a
tightly connected clique, with any two reviewers having reviewed at least 115
and at most 164 apps in common.

\begin{table}[t]
\setlength{\tabcolsep}{.35em}
%\vspace{-10pt}
\centering
\textsf{
\begin{tabular}{l | r r | r}
\toprule
\textbf{Strategy} & \textbf{FPR}\% & \textbf{FNR}\% & \textbf{Accuracy}\%\\
\midrule
%Bagging & 96.70 & 4.02 & 3.30 & 96.35\\
FairPlay/DT & 4.02 & 4.25 & 95.86\\
FairPlay/MLP & 4.52 & 4.72 & 95.37\\
FairPlay/RF & \textbf{1.51} & 6.13 & \textbf{96.11}\\
\midrule
Sarma et al.~\cite{SLGPNRM12}/SVM & 65.32 & 24.47 & 55.23\\
\bottomrule
\end{tabular}
}
\vspace{-5pt}
\caption{
FairPlay classification results ($10$-fold cross validation) of gold standard
malware (positive) and benign apps, significantly outperforming Sarma et al.~\cite{SLGPNRM12}.
FairPlay's RF achieves 96.11\% accuracy at 1.51\% FPR.}
\vspace{-15pt}
\label{table:fp:malware}
\end{table}

\noindent
{\bf Malware Detection Accuracy}.
We have used Sarma et al.~\cite{SLGPNRM12}'s solution as a baseline to evaluate
the ability of FairPlay to accurately detect malware. We computed Sarma et
al.~\cite{SLGPNRM12}'s RCP and RPCP indicators (see
$\S$~\ref{sec:related:malware}) using the longitudinal app dataset. We used the
SVM based variant of Sarma et al.~\cite{SLGPNRM12}, which performs best.
Table~\ref{table:fp:fraudulent} shows $10$-cross validation results over the
malware and benign gold standard sets. FairPlay significantly outperforms Sarma
et al.~\cite{SLGPNRM12}'s solution, with an accuracy that consistently exceeds
$95$\%. 
Random Forest has the smallest FPR of $1.51$\% and the highest accuracy of $96.11\%$.
% and achieves an EER of $4$\% and an AUC of $0.986$. 
This is surprising: most FairPlay
features are meant to identify search rank fraud, yet they \textit{also} accurately
identify malware. 

%\ratul{for 60-40 train vs. test and malware classification, accuracy of the
%classifiers: DT(96.34\%), RF(95.73\%), MLP (95.73\%) and the same ratio with
%fraud classification, accuracy of the classifiers: DT(95.59\%), RF(97.48\%),
%MLP (96.85\%). For 80-20 train vs. test and malware classification, accuracy of
%the classifiers: DT(96.34\%), RF(96.34\%), MLP (95.12\%) and the same ratio
%with fraud classification, accuracy of the classifiers: DT(97.5\%), RF(97.5\%),
%MLP (97.5\%)}
%\bogdan{It seems DT is more efficient (96.34\%) for 60-to-40 than for
%10-fold cross validation}

%
% placing this data in such a table is incorrect
%
%\begin{table}[h]
%\centering
%\textsf{
%\begin{tabular}{l l l l l}
%\toprule
%\textbf{Classifier} & \textbf{TPR(\%)} & \textbf{FNR(\%)} & \textbf{Acc(\%)}\\
%\midrule
%%Bagging & 69.81 & 30.19 & 69.81\\
%RF & 72.16 & 27.83 & 72.17\\
%DT & 75.94 & 24.06 & 75.94\\
%MLP & 60.85 & 39.15 & 60.85\\
%\bottomrule
%\end{tabular}
%}
%\vspace{-5pt}
%\caption{Search rank fraud among malware apps. 10-fold cross validation results
%of classification of the gold standard malware apps when FairPlay is trained on
%all the fraudulent and legitimate apps. Decision Tree flaggs almost 76\% of the
%malware as fraudulent. \bogdan{Are we also updating these results, when removing the
%topD feature?}}
%\vspace{-15pt}
%\label{table:fp:malware:fraud}
%\end{table}

\noindent
{\bf Is Malware Involved in Fraud?}
We conjectured that the above result is due in part to malware apps being
involved in search rank fraud. To verify this, we have trained FairPlay on the
gold standard benign and fraudulent app datasets, then we have tested it on the
gold standard malware dataset.  MLP is the most conservative algorithm,
discovering $60.85$\% of malware as fraud participants.  Random Forest
discovers $72.15$\%, and Decision Tree flags $75.94$\% of the malware as
fraudulent. This result confirms our conjecture and shows that search rank
fraud detection can be an important addition to mobile malware detection
efforts.

\subsection{FairPlay on the Field.}
\label{sec:evaluation:field}

%
% for journal
%
%\begin{figure}
%\centering
%\vspace{-5pt}
%\includegraphics[width=0.45\textwidth]{graphs/wild/strong.signal.eps}
%\caption{Distribution of the number of strong signal words in the wild
%apps.
%\vspace{-15pt}
%\label{fig:strong:signal}
%\end{figure}

We have also evaluated FairPlay on non ``gold standard'' apps. For this, we
have collected a set of apps, as follows. First, we selected $8$ app
categories: Arcade, Entertainment, Photography, Simulation, Racing, Sports,
Lifestyle, Casual. We have selected the $6,300$ apps from the longitudinal
dataset of the $87$K apps, that belong to one of these $8$ categories, and that
have more than $10$ reviews. From these $6,300$ apps, we randomly selected
$200$ apps per category, for a total of $1,600$ apps. We have then collected
the data of all their $50,643$ reviewers (not unique) including the ids of all
the $166,407$ apps they reviewed.

We trained FairPlay with Random Forest (best performing on previous
experiments) on all the gold standard benign and fraudulent apps. We have
then run FairPlay on the $1,600$ apps, and identified $372$ apps (23\%) as
fraudulent.  The Racing and Arcade categories have the highest fraud densities:
$34$\% and $36$\% of their apps were flagged as fraudulent.

\noindent
{\bf Intuition}.
During the $10$-fold cross validation of FairPlay for the gold standard
fraudulent and benign sets, the top most impactful features for the
Decision Tree classifier were (i) the percentage of nodes that belong to the
largest pseudo clique, (ii) the percentage of nodes that belong to at least one
pseudo clique, (iii) the percentage of reviews that contain fraud indicator
words, and (iv) the number of pseudo clique with $\theta \ge 3$.

We use these features to offer an intuition for the surprisingly high fraud
percentage ($23$\% of $1,600$ apps). Figure~\ref{fig:clique:count} shows that
$93.3$\% of the $372$ apps have at least $1$ pseudo clique of $\theta \ge 3$,
nearly $71$\% have at least $3$ pseudo cliques, and a single app can have up to
$23$ pseudo cliques.  Figure~\ref{fig:clique:members} shows that the pseudo
cliques are large and encompass many of the reviews of the apps: $55$\% of the
$372$ apps have at least $33$\% of their reviewers involved in a pseudo clique,
while nearly $51$\% of the apps have a single pseudo clique containing $33$\%
of their reviewers.  While not plotted here due to space constraints, we note
that around $75$\% of the $372$ fraudulent apps have at least $20$ fraud
indicator words in their reviews.

\vspace{-5pt}

\subsection{Coercive Campaign Apps.}
\label{sec:evaluation:coercive}

Upon close inspection of apps flagged as fraudulent by FairPlay, we identified
apps perpetrating a new attack type. The apps, which we call {\it coercive
campaign apps}, harass the user to either (i) write a positive review for the
app, or (ii) install and write a positive review for other apps (often of the
same developer). In return, the app rewards the user by, e.g., removing ads,
providing more features, unlocking the next game level, boosting the
user's game level or awarding game points.

We found evidence of coercive campaign apps from users complaining through
reviews, e.g., ``I only rated it because i didn't want it to pop up while i am
playing'', or ``Could not even play one level before i had to rate it [...]
they actually are telling me to rate the app $5$ stars''.

We leveraged this evidence to identify more coercive campaign apps from the
longitudinal app set. Specifically, we have first manually selected a list
of potential keywords indicating coercive apps (e.g., ``rate'', ``download'',
``ads'').  We then searched all the $2,850,705$ reviews of the $87$K
apps and found around $82$K reviews that contain at least one of these
keywords. Due to time constraints, we then randomly selected $3,000$ reviews
from this set, that are not flagged as fraudulent by FairPlay's RF module. Upon
manual inspection, we identified $118$ reviews that report coercive apps, and
$48$ apps that have received at least $2$ such reviews.  We leave a more
thorough investigation of this phenomenon for future work.

\vspace{-5pt}

\section{Conclusions}

We have introduced FairPlay, a system to detect both fraudulent and malware
Google Play apps.  Our experiments on a newly contributed longitudinal app
dataset, have shown that a high percentage of malware is involved in search
rank fraud; both are accurately identified by FairPlay. In addition, we showed
FairPlay's ability to discover hundreds of apps that evade Google Play's
detection technology, including a new type of \mbox{coercive} fraud attack.

\section{Acknowledgments}

This research was supported in part by NSF grants 1527153 and 1526254, and DoD
W911NF-13-1-0142.

\vspace{-10pt}
\bibliographystyle{unsrt}
\bibliography{anonymous,bogdan,crowdsource,foursquare,graph,location,osn,sybil,sfe,social.fraud,crypto,malware,ml,privacy,nlp,reviews,watchyt,polo}

\end{document}